\documentclass[10pt,oneside,final]{IEEEtran}

\usepackage{amssymb}
\usepackage{amsmath}

\usepackage{graphicx} % 插入图片
\usepackage{subcaption} % 子图支持
\usepackage{color}
\usepackage{multicol}

\usepackage{bm}
\usepackage{latexsym}
\usepackage{stfloats}
\usepackage{float}
\usepackage{exscale}
\usepackage{relsize}
\usepackage{cite}
\usepackage{cases}
\usepackage{algorithm}
\usepackage{algorithmicx}
\usepackage{algpseudocode}
\usepackage{epsfig}
\usepackage{epstopdf}
\usepackage{breqn}
\usepackage{enumerate}
\usepackage{multirow}
\usepackage[utf8]{inputenc}

\usepackage{mathtools}

\usepackage{setspace}
\usepackage{url}
\usepackage{makecell}
\usepackage{diagbox}
\usepackage{stmaryrd}

\usepackage{xcolor}
\usepackage{xpatch}

\usepackage[colorlinks,
linkcolor=blue,
urlcolor=black,
anchorcolor=blue,
citecolor=blue]{hyperref}
\usepackage{accents} % 为了使用\accentset给变量添加单字符修饰符

\algrenewcommand{\algorithmicrequire}{\textbf{Input:}}  % Use Input in the format of Algorithm  
\algrenewcommand{\algorithmicensure}{\textbf{Output:}} % Use Output in the format of Algorithm  

\begin{document}

	\title{Subspace Fusion Sensing for Cooperative ISAC}
	
	\author{Yining Xu, Cunhua Pan, \IEEEmembership{Senior Member, IEEE}, Jun Tang, Hong Ren, and Jiangzhou Wang, \IEEEmembership{Fellow, IEEE}
		\thanks{
			The authors are with National Mobile Communications Research Laboratory, Southeast University, Nanjing 210096, China (e-mail:  yining.xu, cpan, 230268160, hren, j.wang@seu.edu.cn). \textit{Corresponding author: Cunhua Pan}}
		%\vspace{-0.5cm}
	}
	%
	%\IEEEoverridecommandlockouts
	
	%%%%%%%%%%%%%%%%%%wrs%%%%%%%%%%%%%%%%%%%%%%%%%%%%%%%%%%%%%%%
	%
	%   \author{
	% 	           Author 1, Author 2, Author 3, and Author 4
	% 	  %\vspace{-0.9cm}
	% 	  %\vspace{-0.6cm}
	% 	  \vspace{-0.3cm}
		
	% 	  \thanks{Message of Author2}
	% 	  \thanks{This work was supported in part by}
	% 	}
	%%%%%%%%%%%%%%%%%%%%%%%%%%%%%%%%20201126%%%%%%%%%%%%%%%%%%%%
	\maketitle	
	\newtheorem{lemma}{Lemma}
	\newtheorem{theorem}{Theorem}
	\newtheorem{remark}{Remark}
	\newtheorem{corollary}{Corollary}
	\newtheorem{proposition}{Proposition}
	%\allowdisplaybreaks[4]
	
	% 自定义方式实现Axmath中的\mathcalbf{}指令，以打出加粗的花体字母
	\newcommand{\mathcalbf}[1]{\bm{\mathcal{#1}}}

	%%%%%%%%%%%%%%%%%%wrs%%%%%%%%%%%%%%%%%%%%%%%%%%%%%%%%%%%%%%%
	%
	%%%%%%%%%%%%%%%%%%%%%%%%%%%%%%%%20201126%%%%%%%%%%%%%%%%%%%%
	\vspace{-0.8cm}
	\begin{abstract}
		% The rise of the Low-Altitude Economy (LAE) necessitates high-precision 3D localization and 3D velocity estimation. However, traditional Integrated Sensing and Communication (ISAC) techniques are often insufficient for this demanding task. 
		This paper proposes a subspace fusion sensing algorithm for cooperative integrated sensing and communication (ISAC). First, we stack the received signals from access points (APs) into a third-order tensor and construct the equivalent virtual antenna (EVA) array via tensor unfolding. Then, a data association-free subspace-based fusion sensing algorithm is developed utilizing the EVA arrays from distributed APs. A derivation of Cram\'er-Rao lower bound (CRLB) is also presented. Finally, simulation results validate the effectiveness of the proposed algorithm compared to traditional techniques.

		\begin{IEEEkeywords}
		Cooperative integrated communication and sensing, tensor, data fusion.
		\end{IEEEkeywords}
		
	\end{abstract}

	%%%%%%%%%%%%%%%%%%wrs%%%%%%%%%%%%%%%%%%%%%%%%%%%%%%%%%%%%%%%
	%
	% \section{}
	%%%%%%%%%%%%%%%%%%%%%%%%%%%%%%%%20201126%%%%%%%%%%%%%%%%%%%%
	
	%%%%%%%%%%%%%%%%%%wrs%%%%%%%%%%%%%%%%%%%%%%%%%%%%%%%%%%%%%%%
	%
	%\begin{figure}
	%\centering
	%\includegraphics[scale=0.67]{systemmodel}
	%\caption{Illustration of the IRS-aided FD two-way communication between a MIMO BS and $K$ SISO users.}
	%\label{Figsysmodel}
	%\end{figure}
	%%%%%%%%%%%%%%%%%%%%%%%%%%%%%%%%20201126%%%%%%%%%%%%%%%%%%%%
	
	%%%%%%%%%%%%%%%%%%wrs%%%%%%%%%%%%%%%%%%%%%%%%%%%%%%%%%%%%%%%
	%
	% \section{}
	%%%%%%%%%%%%%%%%%%%%%%%%%%%%%%%%20201126%%%%%%%%%%%%%
	\vspace{-0.4cm}
	\section{Introduction}
% The low-altitude economy (LAE), driven by the rapid proliferation of UAVs and new forms of aerial mobility, serves as a catalyst for emerging sectors such as aerial logistics and emergency response.
Integrated sensing and communication (ISAC) has been envisioned as a pivotal technology for the sixth-generation (6G) mobile networks, supporting emerging applications such as the low-altitude economy and autonomous driving \cite{9737357}. However, conventional ISAC systems typically operate in a single-node configuration, where an individual base station (BS) performs sensing tasks independently via signal processing techniques such as tensor decomposition \cite{10403776} and compressive sensing \cite{zhang2025modular}. \textcolor{blue}{To improve the sensing performance, array-side design flexibility has been studied in single-node ISAC\cite{10870338}}.  \textcolor{blue}{Nevertheless, the above techniques still cannot overcome the inherent physical limitations of single-node ISAC}. 
% This configuration faces inherent physical limitations. 
In particular, the severe round-trip path loss and the limited power of BSs significantly degrade sensing performance for distant targets, which is exacerbated when line-of-sight (LoS) paths are obstructed. 

To mitigate these issues, cooperative ISAC has emerged as a promising paradigm, leveraging existing communication networks for multi-node cooperative sensing. Recently, several cooperative sensing schemes have been proposed. Tang et al. proposed a multi-target sensing algorithm based on tensor decomposition \cite{JunTang}. Zhang et al. developed a single-antenna sensing scheme utilizing maximum likelihood estimation (MLE) \cite{zhenkunzhang}. Liu et al. utilized multipath components to establish virtual BSs to sense a single target \cite{10978578}. Liu et al. proposed a Bayesian probability fusion approach for cooperative ISAC \cite{liu2025bayesianprobabilityfusionmultiap}. However, most existing works rely on two-stage framework including complex data association, which may not fully exploit the cooperative sensing gain. In this paper, different from the aforementioned works, we propose a novel cooperative sensing algorithm based on subspace fusion. The main contributions are summarized as follows:
\begin{itemize}
    \item We construct equivalent virtual antenna (EVA) arrays via tensor unfolding to synthesize a larger sensing aperture.
	% We propose an equivalent virtual antenna (EVA) array via tensor unfolding to synthesize a larger sensing aperture.
    \item We propose an association-free subspace fusion algorithm for high-precision sensing.
    \item We derive the Cramér-Rao lower bound (CRLB) and verify its superiority over traditional techniques.
\end{itemize}
\vspace{-3ex}
\section{System model}
\begin{figure}[!t]
		\centering
		\vspace{-6ex} % 向上移动
		\includegraphics[width=0.65 \linewidth]{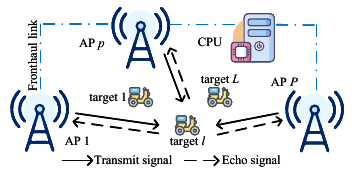}
		\caption{cooperative ISAC system.}
		\label{fig:system_model}
	\end{figure}
The cooperative ISAC system is illustrated in Fig. \ref{fig:system_model}. Consider a cellular network comprising $P$ access points (APs) to sense $L$ targets. Each AP utilizes a uniform linear array (ULA) with $M$ antenna elements and operates in the monostatic mode. \textcolor{blue}{Considering that practical cooperative ISAC systems only involve a limited number of APs, frequency-division-based AP-signal separability is acceptable in many scenarios \cite{JunTang} \cite{shi2022device}. Based on this assumption, the APs broadcast 5G New Radio (5G NR) orthogonal frequency-division multiplexing (OFDM) signals across $P$ distinct frequency bands, and the central control unit (CPU) aggregates the sensing data for joint estimation, which allows us to focus on the proposed subspace fusion sensing framework while keeping the system model tractable.}
% Consistent with \cite{JunTang}, the APs broadcast 5G New Radio (5G NR) orthogonal frequency-division multiplexing (OFDM) signals across $P$ distinct frequency bands
% % \footnote{\textcolor{blue}{Zadoff-Chu (ZC) sequences can also be considered for distinguishing signals from different APs, as discussed in \cite{liu2025bayesianprobabilityfusionmultiap}.}}
%  and the central control unit (CPU) aggregates the sensing data for joint estimation. \textcolor{blue}{Since practical cooperative ISAC systems usually involve a limited number of cooperating APs, this assumption is acceptable in many scenarios and allows us to focus on the proposed subspace fusion sensing framework while keeping the system model tractable.}.
The transmitted signal of the $p$th AP is given by
% {\textcolor{blue}{Zadoff-Chu (ZC) sequences can also ensure AP-signal orthogonality as an alternative to  frequency-division schemes, as discussed in \cite{liu2025bayesianprobabilityfusionmultiap}.}}
% \footnote{\textcolor{blue}{Zadoff-Chu (ZC) sequences can alternatively distinguish AP signals when distinct bands are unavailable \cite{liu2025bayesianprobabilityfusionmultiap}.}}
$
s_{p}\left( t \right) =\sum_{k=1}^K{\sum_{n=1}^N{s_{p,k,n}}e^{j2\pi k\Delta ft}}\cdot r\left( t-nT_{s} \right),
$
% \begin{align}
% 		s_{p}\left( t \right) =\sum_{k=1}^K{\sum_{n=1}^N{s_{p,k,n}}e^{j2\pi k\Delta ft}}\cdot r\left( t-nT_{s} \right),
% \end{align}  
where $K$ and $N$ are respectively the number of subcarriers and OFDM symbols, $s_{p,k,n}$ is the modulated symbols allocated in the $k$th subcarrier, $n$th OFDM symbol at the $p$th AP, $T_{\mathrm{s}}$ is the OFDM symbol period (including the cyclic prefix), $r(t)$ is the pulse-shaping filter function and $\Delta f$ is the subcarrier spacing (SCS). 
% Then, the transmitted frequency domain symbol vector can be denoted as $\mathbf{e}_p=\mathbf{f}_{p} s_{p}$, where $\mathbf{f}_{p} \in \mathbb{C}^{M\times 1}$ denotes the beamforming vector at AP $p$. 
Without loss of generality, we assume that $|s_{p,k,n}|^2 = 1, \forall p, k, n$. The frequency domain signal vector is given by
\begin{align}
	\mathbf{y}_{p,k,n}=\mathbf{H}_{p,k,n}\mathbf{f}_{p} s_{p,k,n}+\mathbf{n}_{p,k,n},
	\label{equation:receive_signal}
\end{align}
where $\mathbf{f}_{p} \in \mathbb{C}^{M\times 1}$ denotes the beamforming vector at the $p$th AP, $\mathbf{H}_{p,k,n}\in \mathbb{C}^{M\times M}$ denotes the discrete frequency-domain monostatic channel at the $p$th AP and $\mathbf{n}_{p,k,n}\in\mathbb{C}^{M\times1}$ denotes the additive white Gaussian noise (AWGN) vector. To eliminate the sensing symbols, the received signal is divided by the transmitted symbols, i.e.,
\begin{align}
	\hat{\mathbf{y}}_{p,k,n}=\mathbf{y}_{p,k,n} / s_{p,k,n}= \mathbf{H}_{p,k,n}\mathbf{f}_{p}+\hat{\mathbf{n}}_{p,k,n},
	\label{equation:bistatic_y} 
\end{align}
% \begin{align}
where $\hat{\mathbf{n}}_{p,k,n}$ denotes the equivalent noise vector.  Then, the time and delay domain sensing channel is modeled as
\begin{align}
	\mathbf{H}_{p}(t,\tau) = \sum_{l=1}^{L}  
			\alpha_{p,l} \delta(\tau-\tau_{p,l}) \mathbf{a}_{p}\left( \theta_{p,l} \right)  
			\cdot \mathbf{a}_{p}^{H}\left( \theta_{p,l} \right)e^{j2\pi f_{p,l}^{d}t}, 
	\end{align}
	 where $\alpha_{p,l}$, $\tau_{p,l} = 2d_{p,l}/c$ and $f_{p,l}^{d} = 2v_{p,l}/{\lambda}$ are respectively the complex channel gains, delay and Doppler shift. $d_{p,l}$, $v_{p,l}$ are respectively the range and radial velocity. The symbols  $c$ and $\lambda$ are respectively the speed of light and wavelength. The symbol $\theta_{p,l}$ is the angle of \textcolor{blue}{arrival} (AoA). For brevity, we define the virtual angle as $\psi_{p,l} \triangleq \cos(\theta_{p,l})$. Thus, the steering vector  is rewritten as 
	%  \begin{align}
	%  \mathbf{a}_{p}(\psi_{p,l}) = \left[ 1, \cdots, e^{j2\pi (M-1) \frac{d}{\lambda} \psi_{j,l}} \right]^T.
	%  \label{equation:ap}
	%  \end{align}
	% \begin{align}
	% \mathbf{a}_{p}(\psi_{p,l}) = \left[ 1, \cdots, e^{j2\pi (M-1) \frac{d}{\lambda} \psi_{j,l}} \right]^T.
	%  \label{equation:ap}
	% \end{align}
	 $
	 \mathbf{a}_{p}(\psi_{p,l}) = \left[ 1, \cdots, e^{j2\pi (M-1) \frac{d}{\lambda} \psi_{p,l}} \right]^T.
	 \label{equation:ap}
	$
	% where $\theta_{j,k}$ and $\phi_{j,k}$ denote the elevation and azimuth angles of arrival/departure (AoA/DOA)\footnote{ AOA and DOA equals in the monostatic ISAC} and  at BS $j$ for the target $k$, respectively. To simple the the notation, we define the virtual angles as $\Theta_{j,k} \triangleq \sin(\theta_{j,k}) \cos(\phi_{j,k})$ and $\Phi_{j,k} \triangleq \sin(\theta_{j,k})\cos(\phi_{j,k})$, respectively.
Then, following the Fourier transform (FT) over the delay and sampling at the $n$th OFDM symbol, the discrete frequency domain channel is given by 
	\begin{align}
		\mathbf{H}_{p,k,n} = \sum_{l=1}^{L} 
			\alpha_{p,l} \mathbf{a}_{p}\left( \psi_{p,l} \right)  \mathbf{a}_{p}^{H}\left( \psi_{p,l} \right)
			 e^{-j2\pi k\Delta f\tau_{p,l}} e^{j2\pi f_{p,l}^{d} nT_{s}}.
	\label{equation:channel_model}
			\end{align}

%\vspace{-6ex}
\section{Subspace fusion sensing algorithm}
\begin{figure}[!t]
		\centering
		\vspace{-1ex} % 向上移动
		\includegraphics[width=0.9 \linewidth]{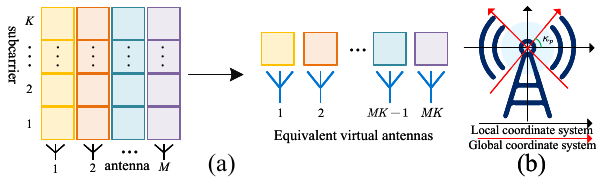}
		\caption{Illustration of (a) EVA array construction and (b) local and global coordinate systems at the AP.}
		\label{fig:EVA}
	\end{figure}

%\vspace{-2ex}
By substituting \eqref{equation:channel_model} into \eqref{equation:receive_signal}, the received signal at the $p$th AP is written as a third-order tensor $\mathcalbf{Y}_p\in \mathbb{C}^{M \times N \times K}$, i.e.,
\begin{equation}
    \mathcalbf{Y}_p = \sum_{l=1}^{L} \beta_{p,l} \mathbf{a}_p(\psi_{p,l}) \circ \mathbf{o}_p({f}_{p,l}^d) \circ \mathbf{g}_p(\tau_{p,l}) + \mathcalbf{N}_p , 
    \label{eq:tensor_model}
\end{equation}
where $\mathcalbf{N}_p \in \mathbb{C}^{M \times N \times K}$ is the equivalent noise tensor and $\beta_{p,l} =  \alpha_{p,l}\mathbf{a}_p^H(\psi_{p,l})\mathbf{f}_p$, $\mathbf{o}_p(f_{p,l}^d) = \left[  e^{j2\pi T_s f^d_{p,l}}, \dots, e^{j2\pi NT_s f^d_{p,l}} \right]^T \in \mathbb{C}^{N \times 1}$, 
$\mathbf{g}_p(\tau_{p,l}) = \left[ e^{-j2\pi \Delta f \tau_{p,l}}, \dots, e^{-j2\pi K\Delta f \tau_{p,l}} \right]^T \in \mathbb{C}^{K \times 1}$ and $\circ$ denotes the vector outer product. The factor matrices of $\mathcalbf{Y}_p$ are respectively denoted as $\mathbf{A}_p = [\mathbf{a}_p(\psi_{p,1}), \dots, \mathbf{a}_p(\psi_{p,L})] \in \mathbb{C}^{M \times L}$, $\mathbf{O}_p = [\mathbf{o}_p(f_{p,1}^d), \dots, \mathbf{o}_p(f_{p,L}^d)] \in \mathbb{C}^{N \times L}$ and $\mathbf{G}_p = [\mathbf{g}_p(\tau_{p,1}), \dots, \mathbf{g}_p(\tau_{p,L})] \in \mathbb{C}^{K \times L}$. 
For brevity, we omit the noise term in the subsequent derivations. By taking the $n$th slice of the tensor $\mathcal{Y}_p$, we obtain
% \begin{align}
% \mathbf{Y}_{p,n} &= \sum_{l=1}^{L} \gamma_{p,n,l} \cdot \mathbf{a}_p(\psi_{p,l}) \mathbf{g}_p^T(\tau_{p,l}) \nonumber \\
% &= \mathbf{A}_p \mathbf{D}_{p,n} \mathbf{G}_p^T.
% \label{equation:Y_slice}
% \end{align}
% \begin{align}
% \mathbf{Y}_{p,n} &= \sum_{l=1}^{L} \gamma_{p,n,l} \cdot \mathbf{a}_p(\psi_{p,l}) \mathbf{g}_p^T(\tau_{p,l}) \nonumber \\
% &= \mathbf{A}_p \mathbf{D}_{p,n} \mathbf{G}_p^T.
% \label{equation:Y_slice}
% \end{align}
\begin{align}
\mathbf{Y}_{p,n} = \sum_{l=1}^{L} \gamma_{p,n,l} \cdot \mathbf{a}_p(\psi_{p,l}) \mathbf{g}_p^T(\tau_{p,l}) = \mathbf{A}_p \mathbf{D}_{p,n} \mathbf{G}_p^T.
\label{equation:Y_slice}
\end{align}
where $\gamma_{p,n,l} = \beta_{p,l} e^{j2\pi f_{p,l}^d nT_s}$,
% We rewrite \eqref{equation:Y_slice} in a matrix form as
% \begin{align}
% \mathbf{Y}_{p,n} = \mathbf{A}_p \mathbf{D}_{p,n} \mathbf{G}_p^T + \mathbf{N}_{p,n}.
% \end{align}
$\mathbf{D}_{p,n} = \text{diag}\left( \mathbf{s}_{p,n} \right)$ and $\mathbf{s}_{p,n} = [\gamma_{p,n,1}, \dots, \gamma_{p,n,L}]^T$. By performing the $\text{vec}$ operator, we have
% \begin{align}\mathbf{y}_{p,n} &= \text{vec}(\mathbf{Y}_{p,n}) = \left( (\mathbf{G}_p^T)^T \otimes \mathbf{A}_p \right) \text{vec}(\mathbf{D}_{p,n}) + \text{vec}(\mathbf{N}_{p,n}) \nonumber\\ &=
% 	\left( \mathbf{G}_p \odot \mathbf{A}_p \right) \mathbf{s}_{p,n} + \mathbf{n}_{p,n} \triangleq \mathbf{\Pi }_p \mathbf{s}_{p,n} + \mathbf{n}_{p,n},
% \label{equation:vec}
% \end{align}
\begin{align}\mathbf{y}_{p,n} &= \text{vec}(\mathbf{Y}_{p,n}) = \left( (\mathbf{G}_p^T)^T \otimes \mathbf{A}_p \right) \text{vec}(\mathbf{D}_{p,n})  \nonumber\\ &=
	\left( \mathbf{G}_p \odot \mathbf{A}_p \right) \mathbf{s}_{p,n}  \triangleq \mathbf{\Pi }_p \mathbf{s}_{p,n} ,
\label{equation:vec}
\end{align}
where $\otimes$ and $\odot$ are respectively the Kronecker product and Khatri-Rao product. The unknown parameters in \eqref{equation:vec} can be expressed as $
    \boldsymbol{\xi} = \left[ \text{Re}(\mathbf{s})^T, \text{Im}(\mathbf{s})^T, \mathbf{e}^T \right]^T$
where $\mathbf{s} = [\mathbf{s}_1^T, \dots, \mathbf{s}_P^T]^T$, $\mathbf{s}_p = [\mathbf{s}_{p,1}^T, \dots, \mathbf{s}_{p,N}^T]^T$, $\mathbf{e} = [(\mathbf{p}_1^{t})^T,  \dots, (\mathbf{p}_L^{t,T})^T]^T$ and $\mathbf{p}_l^t=[x_l^t,y_l^t]^T$ is the position of the $l$th target. Thus, the likelihood function for $\boldsymbol{\xi}$ is given by
\begin{align}
f(\mathbf{y}|\boldsymbol{\xi}) =  \frac{1}{(\pi \sigma_n^2)^{MKNP}} e^{ -\frac{1}{\sigma_n^2} \sum_{p,n} \|\mathbf{y}_{p,n} - \mathbf{\Pi }_p \mathbf{s}_{p,n}\|^2  },
\label{eq:likelihood_function}
\end{align}
where $\mathbf{y} = [\mathbf{y}_1^T, \dots, \mathbf{y}_P^T]^T$, $\mathbf{y}_p = \text{vec}(\mathcalbf{Y}_p)$, $\sigma_n^2$ is the variance of the AWGN and $\left\| \cdot \right\|$ denotes the L2-norm. According to \eqref{eq:likelihood_function}, the log-likelihood function can be expressed as
\begin{align}
f_{ln}(\mathbf{y}|\boldsymbol{\xi}) = - MKNP \ln\pi \sigma_n^2 \nonumber -  \frac{1}{\sigma_n^2} \sum_{p,n} \|\mathbf{y}_{p,n} - \mathbf{\Pi }_p \mathbf{s}_{p,n}\|^2.
\label{equation:log_like}
\end{align}
We define $S_{p,n}(\xi) = \sum_{p,n} \|\mathbf{y}_{p,n} - \mathbf{\Pi }_p \mathbf{s}_{p,n}\|^2$. Differentiating  $f_{ln}(\mathbf{y}|\boldsymbol{\xi})$ with respect to $\sigma^2_n$ and  letting it equal to 0, we have
\begin{align}
\frac{\partial f_{ln}}{\partial \sigma_n^2} = -\frac{MKNP}{\sigma_n^2} + \frac{S_{p,n}(\xi)}{(\sigma_n^2)^2} = 0.
\end{align}
Thus, the noise power can be estimated as $\hat{\sigma}_n^2 = \frac{1}{MKNP} S_{p,n}(\xi)$.
By substituting $\hat{\sigma}_n^2$ into \eqref{equation:log_like}, we have
\begin{align}
f_{ln}(\mathbf{y}|\boldsymbol{\xi}) &= - MKNP \ln\hat{\sigma}_n^2 - \frac{S_{p,n}(\xi)}{\hat{\sigma}_n^2}
\nonumber \propto -\ln(S_{p,n}(\xi)),
\end{align}
where $\propto$ denotes proportionality. Thus, we can obtain the optimal $\boldsymbol{\xi}$ by minimizing the following cost function, i.e.,
\begin{align}
\hat{\boldsymbol{\xi}} = \arg \min_{\boldsymbol{\xi}} \ln(S_{p,n}(\xi)).
\label{equation:search_xi}
\end{align}
Problem \eqref{equation:search_xi} presents a high-dimensional, nonlinear optimization challenge characterized by a complex cost function. Consequently, deriving a closed-form analytical expression for $\hat{\xi}$ is mathematically intractable. Although exhaustive grid search methods could theoretically locate the global optimum, the associated computational complexity is prohibitively high, rendering such an approach impractical for practical applications. Therefore, we simplify this problem from a subspace perspective to reduce the dimension of searching. A concept named EVA array is proposed in Fig. \ref{fig:EVA}. \textcolor{blue}{The EVA array is a spatial-frequency equivalent representation and does not create additional antenna locations in physical space.} 
% \ref{fig:EVA}a \footnote{\textcolor{blue}{It is noted that the EVA array is a spatial-frequency equivalent representation rather than a structure that creates additional antenna locations in physical space.}}.
Specifically, we can fully exploit the degrees of freedom in both the spatial and frequency domains to achieve enhanced sensing performance by treating the signals received across $M$ physical antennas and $K$ subcarriers as observations from a virtual antenna array with $MK$ equivalent elements. To achieve this idea, we first perform the mode-2 unfolding of  $\mathcalbf{Y}_p$, given by
\begin{equation}
\mathbf{Y}_{(2)} = \left( \mathbf{G}_p \odot \mathbf{A}_p \right) \mathbf{O}_p^{T}.
\label{equation:Y_unfolding}
\end{equation}
The mode-2 unfolding operation in \eqref{equation:Y_unfolding} stacks the spatial dimension and the frequency dimension. Thus, each EVA corresponds to a specific combination of a physical antenna index and a subcarrier index in the EVA array. Then, the covariance matrix of \eqref{equation:Y_unfolding} can be estimated as
\begin{align}
	\hat{\mathbf{R}}_p = \frac{1}{N} \mathbf{Y}_{(2)} \mathbf{Y}_{(2)}^H  = \frac{1}{N}\left( \mathbf{G}_p \odot \mathbf{A}_p \right) \hat{\mathbf{R}}_{o,p} \left( \mathbf{G}_p \odot \mathbf{A}_p \right)^H ,
	\label{equation:variance}	
	\end{align}
	where $\hat{\mathbf{R}}_{o,p} =  \mathbf{O}_p^T\mathbf{O}_p^*$. By performing the eigenvalue decomposition (EVD) of $\hat{\mathbf{R}}_p$, we have
	\begin{equation}
\hat{\mathbf{R}}_p = \sum_{i=0}^{L} \lambda_i \boldsymbol{\mu}_i \boldsymbol{\mu}_i^H + \sum_{i=L+1}^{MK-1} \lambda_i \boldsymbol{\mu}_i \boldsymbol{\mu}_i^H,
\label{equation:noise_space}
\end{equation}
where $\lambda_i$ and $\boldsymbol{\mu}_i$ are the $i$th eigenvalue and eigenvector of $\hat{\mathbf{R}}_p$, respectively. The eigenvalues are sorted in non-decreasing order. Thus, $\mathbf{U}_{p,a} = [ \boldsymbol{\mu}_0, \dots, \boldsymbol{\mu}_L]$ forms the signal subspace and $\mathbf{U}_{p,w} = [ \boldsymbol{\mu}_{L+1}, \dots, \boldsymbol{\mu}_{MK-1}]$ forms the noise subspace. By utilizing the EVA array, the dimension of the noise subspace $\mathbf{U}_{p,w}$ is expanded to $MK-L$. Based on the EVA array structure in \eqref{equation:Y_unfolding}, for any given candidate position $\mathbf{p}^t$, the steering vector of the EVA array is constructed as 
\begin{align}
\bm{\omega}(\mathbf{p}^t) = \mathbf{o}(\tau(\mathbf{p}^t)) \otimes \mathbf{a}(\psi(\mathbf{p}^t)),
\label{equation:eva_steering_vector}
\end{align}
where $\mathbf{o}(\tau(\mathbf{p}^t))$ and $\mathbf{a}(\psi(\mathbf{p}^t))$ are defined in \eqref{eq:tensor_model}, which captures the phase variation across subcarriers and antennas respectively. This construction maps the range and angle information of the target into a unified high-dimensional subspace, eliminating data association and enabling high-precision sensing via orthogonality. The candidate delay $\tau(\mathbf{p}^t)$ and angle $\psi(\mathbf{p}^t)$ are respectively expressed as 
\begin{align}
\tau \left( \mathbf{p}^t \right) = \frac{\left\| \mathbf{T}^{-1}\left( \kappa _p \right) \mathbf{p}^t-\mathbf{T}^{-1}\left( \kappa _p \right) \mathbf{p}_{p}^{a} \right\|}{2c},
\label{equation:tau_p}\\
\left[ \begin{array}{c}
	\psi \left( \mathbf{p}^t \right)\\
	\sin \left(   \psi^{-1} \left( \mathbf{p}^t \right)  \right)
	\\
\end{array} \right] = \frac{\mathbf{T}^{-1}\left( \kappa _p \right) \mathbf{p}^t-\mathbf{T}^{-1}\left( \kappa _p \right) \mathbf{p}_{p}^{a}}{\left\| \mathbf{T}^{-1}\left( \kappa _p \right) \mathbf{p}^t-\mathbf{T}^{-1}\left( \kappa _p \right) \mathbf{p}_{p}^{a} \right\|},
\label{equation:psi_p}
\end{align}
where $\psi^{-1} = \cos^{-1}\left( \psi \left( \mathbf{p}^t \right) \right)$, $\mathbf{p}_{p}^{a}= [x_{p}^{a}, y_{p}^{a}]^T$ is the position of the $p$th AP, $\kappa_p$ is the angle between the local and global coordinate systems at the $p$th AP, as shown in Fig. \ref{fig:EVA}b and $\mathbf{T}(\kappa _p)$ is the transformation matrix from the local coordinate to the global coordinate at the $p$th AP, given by
\begin{align}
\mathbf{T}\left( \kappa _p \right) =\left[ \begin{matrix}
	\cos \left( \kappa _p \right)&		\sin \left( \kappa _p \right)\\
	-\sin \left( \kappa _p \right)&		\cos \left( \kappa _p \right)\\
\end{matrix} \right] .
\label{equation:ratation_matrix}
\end{align}
Combining the observed data of APs, we define the cost function of subspace fusion sensing problem as
\begin{equation}
    \Psi(\mathbf{p}^t) = \frac{1}{\sum_{p=1}^{P} \bm{\omega}^H(\mathbf{p}^t) \mathbf{U}_{p,w} \mathbf{U}_{p,w}^H \bm{\omega}(\mathbf{p}^t)}.
    \label{eq:cost_function}
\end{equation}

The subspace fusion sensing algorithm is shown in Algorithm~\ref{algorithm:subspace}. To find the $L$ peaks of \eqref{eq:cost_function}, we firstly construct a coarse grid set $\mathbb{S}_{coarse}$ including $L_x \times L_y$ points. Then, we can obtain the coarse estimate of the target position by finding the $L$ peaks of \eqref{eq:cost_function}. Finally, a coarse estimate is used as the initial value and solve \eqref{equation:search_xi} with the Quasi-Newton method \cite{gill1972quasi} to obtain the final estimate.

\textbf{Complexity Analysis:}  The mode-2 tensor unfolding, covariance matrix estimation and EVD for $P$ APs require a complexity of $\mathcal{O}(P NM^2K^2 + PM^3K^3)$. Evaluating the cost function over $L_x \times L_y$ grid points yields a complexity of $\mathcal{O}(P L_xL_y M^2K^2)$. The fine search utilizing the Quasi-Newton method for $L$ targets incurs a complexity of $\mathcal{O}(P L I M^2K^2)$, where $I$ is the number of iterations. It is worth noting that our proposed algorithm expands the array aperture by constructing the EVA to achieve higher sensing accuracy, this comes at the cost of increased computational complexity. Therefore, the algorithm is suitable  for resource-constrained scenarios with a limited number of antennas or subcarriers. 
% The minimal sensing resource allocation in communication-centric ISAC systems ensures a manageable computational overhead, making the proposed algorithm highly practical for future cooperative networks.
Since communication-centric ISAC systems allocate only a small fraction of resources for sensing, the actual computational overhead remains manageable, rendering the proposed algorithm practical for future cooperative networks. \textcolor{blue}{In practical implementation, the EVD can be performed locally at each AP with the baseband unit (BBU) or edge processing capability, and only the noise subspace or its compressed representation is forwarded to the CPU for fusion, thus reducing the CPU-side burden and fronthaul latency.}

\begin{algorithm}[h]
	%\vspace{-1ex}
	 % 向上移动
    \caption{Subspace Fusion Sensing Algorithm}
	\small % <--- 在这里添加控制字体的命令（只会影响该算法块内部）
	\label{algorithm:subspace}
    \begin{algorithmic}[1]
        \Statex \textbf{Step 1: Subspace Estimation}
        \For{$p = 1$ to $P$}
            \State Perform mode-2 unfolding of $\mathcalbf{Y}_p$  via \eqref{equation:Y_unfolding}.
            \State Estimate the covariance matrix $\hat{\mathbf{R}}_p$ via \eqref{equation:variance}.
            \State Perform EVD on $\hat{\mathbf{R}}_p$ to obtain the noise subspace $\mathbf{U}_{p,w}$ according to \eqref{equation:noise_space}.
        \EndFor
        
        \Statex \textbf{Step 2: Coarse Search}
        \State Define a coarse grid set $\mathbb{S}_{coarse}$ including $L_x \times L_y$ points.
        \For{each grid point $\mathbf{p}^t \in \mathbb{S}_{coarse}$}
            \State Calculate $\bm{\omega}(\mathbf{p}^t)$ via \eqref{equation:eva_steering_vector}, \eqref{equation:tau_p}, \eqref{equation:psi_p} and \eqref{equation:ratation_matrix}.
            \State Compute $\Psi(\mathbf{p}^t)$ via \eqref{eq:cost_function}.
        \EndFor
        \State Obtain the coarse estimate $\{\mathbf{p}^{t,coarse}_l|l=1,\cdots,L\}$ by finding the $L$ peaks of \eqref{eq:cost_function}.
        
        \Statex \textbf{Step 3: Fine Search}
        %\State Define a fine grid set $\mathbb{S}_{fine}$ around 
		\For{$l = 1$ to $L$}
            \State Use the coarse estimate $\mathbf{p}^{t,coarse}_l$ as the initial value.
			\State Solve $\hat{\mathbf{p}}^{t}_l = \arg \max_{\mathbf{p}^t_l} \Psi(\mathbf{p}^t_l)$ with the Quasi-Newton method and obtain the final estimate $\hat{\mathbf{p}}^{t}_l$.
        \EndFor
    \end{algorithmic}
\end{algorithm}

\vspace{-1ex}
\section{Cram\'er-Rao lower bound}
For the $l$th target, define the unknown parameter vector as $\mathbf{\Xi}_l=\left[ \tau _{1,l},\dots ,\tau _{P,l},\psi_{p,1}^{g} ,\dots ,\psi_{P,l}^{g} \right]^T$. \textcolor{blue}{For the CRLB derivation, all APs and target coordinates are unified into the global coordinate system to eliminate the influence of AP-specific local coordinates, where $\psi^g_{p,l}$ is the corresponding AoA calculated therein.}
%  where $\psi_{p,l}^{g}$ is the virtual angle in the global coordinate.
According to \eqref{eq:tensor_model}, the ISAC echo signal of the $p$th AP  sensing signal can be expressed as
\begin{equation}
    y_{p,l} = \sum_{m,k,n}  \beta_{p,l} e^{j2\pi f_{p,l}^d n T_s} e^{-j2\pi k \Delta f \tau_{p,l}} e^{j2\pi m  \psi_{p,l}^{g}d/\lambda} + n_{p,l},
    \label{eq:echo_signal}
\end{equation}
% where $n_{p}^l \sim \mathcal{CN}(0, \sigma^2)$ and
  Based on \eqref{eq:echo_signal}, the probability density function of $\mathbf{\Xi}_l$ is
\begin{equation}
    q(\mathbf{u}; \mathbf{\Xi}_l) = \frac{1}{(2\pi\sigma_n^2)^{L/2}} e^{ -\frac{1}{2\sigma_n^2} \sum_{l=1}^{L} \left| y_{p,l} - \sum_{m,k,n} s_{p,l,m,k,n} \right|^2 },
    \label{eq:pdf}
\end{equation}
where $\mathbf{u} = [y_{p,l}, \dots, y_{P,l}]^T$ denotes a set of echo signals received by $P$ APs and $s_{p,l,m,k,n} = \beta_{p,l} e^{j2\pi f_{p,l}^d n T_s} e^{-j2\pi k \Delta f \tau_{p,l}} e^{j2\pi m \psi_{p,l}^{g}d / \lambda}.$ The log-likelihood function of $\mathbf{\Xi}_l$ is given by
\begin{equation}
     q_{ln}(\mathbf{u}; \mathbf{\Xi}_l) = -\frac{P}{2} \ln(2\pi\sigma_n^2) - \frac{1}{2\sigma_n^2} \sum_{p=1}^{P} \left| y_{p}^l - \sum_{m,k,n} s_{p,l,m,k,n} \right|^2.
    \label{eq:log_likelihood}
\end{equation}
According to the definition of the Fisher information matrix (FIM), the entries of the matrix are determined by the Hessian of the log-likelihood function. By taking the second-order partial derivatives of \eqref{eq:log_likelihood} with respect to the parameter vector $\mathbf{\Xi}_l$ and computing the expectation, the $(i,j)$th element of the FIM is obtained as
$
\left[ \mathbf{F} \right] _{i,j}=\mathbb{E} \left( -\frac{\partial  q_{ln}\left( \mathbf{u};\mathbf{\Xi}_l \right)}{\partial [\mathbf{\Xi}_l]_i\partial [\mathbf{\Xi }_l]_j} \right) \nonumber =\frac{2}{\sigma _{n}^{2}}\mathrm{Re}\left\{ \left( \frac{\partial \mathbf{s}_l\left( \mathbf{\Xi}_l \right)}{\partial [\mathbf{\Xi}_l]_i} \right) ^H\left( \frac{\partial \mathbf{s}_l\left( \mathbf{\Xi }_l \right)}{\partial [\mathbf{\Xi}_l]_j} \right) \right\} ,
$
% \begin{align}
% 	\left[ \mathbf{F} \right] _{i,j}&=\mathbb{E} \left( -\frac{\partial  q_{ln}\left( \mathbf{u};\mathbf{\Xi}_l \right)}{\partial [\mathbf{\Xi}_l]_i\partial [\mathbf{\Xi }_l]_j} \right) \nonumber \\&=\frac{2}{\sigma _{n}^{2}}\mathrm{Re}\left\{ \left( \frac{\partial \mathbf{s}_l\left( \mathbf{\Xi}_l \right)}{\partial [\mathbf{\Xi}_l]_i} \right) ^H\left( \frac{\partial \mathbf{s}_l\left( \mathbf{\Xi }_l \right)}{\partial [\mathbf{\Xi}_l]_j} \right) \right\} ,
% \end{align}
%$$s_p=\sum_{m,k,n}{\beta _{p}^{u}e^{j2\pi f_{D,l}^{u}kT}e^{-j2\pi n\varDelta f\tau _{l}^{u}}e^{j2\pi m\cos \left( \theta _{l}^{u,g} \right) d/\lambda}}$$
where $\mathbf{s}_l=\left[ s_{1,l},\dots ,s_{P,l} \right]^T $ and $s_{p,l}=\sum_{m,k,n} s_{p,l,m,k,n}$. We firstly calculate the partial derivatives of $s_{p,l}$ with respect to $\tau _{p,l}$ and $\psi_{p,l}^{g} $ as follows
\begin{align}
\frac{\partial s_{p,l}}{\partial \tau _{p,l}}&=\frac{\partial \sum_{m,k,n}{\beta _{p,l}e^{j2\pi f_{p,l}^{d}nT_s}e^{-j2\pi k\varDelta f\tau _{p,l}}e^{j2\pi m\psi_{p,l}^{g} d/\lambda}}}{\partial \tau _{p,l}} \nonumber
\\
&=\sum_{m,k,n,}{-j2\pi n\varDelta fs_{p,l,m,k,n}},
\end{align}
\begin{align}
	\frac{\partial s_{p,l}}{\partial \psi_{p,l}^{g}} &=\frac{\partial \sum_{m,k,n}{\beta _{p,l}e^{j2\pi f_{p,l}^{d}nT_s}e^{-j2\pi k\varDelta f\tau _{p,l}}e^{j2\pi m\psi_{p,l}^{g} d/\lambda}}}{\partial \psi_{p,l}^{g}} \nonumber
\\
&=\sum_{m,k,n,}{j2\pi mds_{p,l,m,k,n}/\lambda}.
\end{align}

According to the above derivations, the FIM consists of four block   
diagonal matrices, given by
$
	\mathbf{F}=\left[ \begin{matrix}
	\mathbf{\Psi }&		\mathbf{\Upsilon }\\
	\mathbf{\Upsilon }&		\mathbf{\Omega }\\
\end{matrix} \right],
$
% \begin{align}
% 	\mathbf{F}=\left[ \begin{matrix}
% 	\mathbf{\Psi }&		\mathbf{\Upsilon }\\
% 	\mathbf{\Upsilon }&		\mathbf{\Omega }\\
% \end{matrix} \right],
% \end{align}
where $\mathbf{\Psi } \in \mathbb{C}^{P\times P}$, $\mathbf{\Omega }\in \mathbb{C}^{P\times P}$ and $\mathbf{\Upsilon }\in \mathbb{C}^{P\times P}$ are diagonal matrices. The diagonal elements of these matrices can be calculated as 
\begin{align}
	\mathbf{\Psi }_{p,p}&=\frac{2}{\sigma _{n}^{2}}\mathrm{Re}\left\{ \left( \frac{\partial s_{p,l}}{\partial \tau _{p,l}} \right) ^*\left( \frac{\partial s_{p,l}}{\partial \tau _{p,l}} \right) \right\} \nonumber
\\
&=\frac{2}{\sigma _{n}^{2}}\sum_{m,k,n}{\left| \beta _{p,l} \right|^2\left( 2\pi n\Delta f \right) ^2} \nonumber
\\
&=\frac{4\pi ^2\left| \beta _{p,l} \right|^2\varDelta f^2MKN\left( N+1 \right) \left( 2N+1 \right)}{3\sigma _{n}^{2}},
\end{align}
\begin{align}
	\mathbf{\Omega }_{p,p}&=\frac{2}{\sigma _{n}^{2}}\mathrm{Re}\left\{ \left( \frac{\partial s_{p,l}}{\partial \psi_{p,l}^{g}} \right) ^*\left( \frac{\partial s_{p,l}}{\partial \psi_{p,l}^{g}} \right) \right\} \nonumber
\\
&=\frac{2}{\sigma _{n}^{2}}\sum_{m,k,n}{\left| \beta _{p,l} \right|^2\left( 2\pi md/\lambda \right) ^2} \nonumber
\\
&=\frac{4\pi ^2\left| \beta _{p,l} \right|^2d^2M\left( M+1 \right) \left( 2M+1 \right) KN}{3\sigma _{n}^{2}\lambda ^2},
\end{align}
\begin{align}
\mathbf{\Upsilon }_{p,p}&=\frac{2}{\sigma _{n}^{2}}\mathrm{Re}\left\{ \left( \frac{\partial s_{p,l}}{\partial \tau _{p,l}} \right) ^*\left( \frac{\partial s_{p,l}}{\partial \psi_{p,l}^{g}} \right) \right\} \nonumber
\\
&=\frac{2}{\sigma _{n}^{2}}\sum_{m,k,n}{\left| \beta _{p,l} \right|^2\left( -2\pi n\Delta f \right)} 2\pi md/\lambda \nonumber
\\
&=\frac{-2\pi ^2\left| \beta _{p,l} \right|^2\Delta fdM\left( M+1 \right) N\left( N+1 \right) K}{ \sigma _{n}^{2}\lambda}.
\end{align}
To derive the CRLB, we need to establish the relationship between $\mathbf{\Xi}_l$ and the target position $\mathbf{p}_{l}^{t} = [x_{l}^{t}, y_{l}^{t}]^T$ using the Jacobian matrix. The geometric relationships are given as follows:
$
\tau _{p,l}= \frac{2\left\| \mathbf{p}_{p}^{a}-\mathbf{p}_{l}^{t} \right\|}{c}, \  \tan \left( \theta _{p,l}^{g} \right)  = \frac{y_{l}^{t}-y_{p}^{a}}{ x_{l}^{t}-x_{p}^{a} }.
$
% \begin{align}
% \tau _{p,l}= \frac{2\left\| \mathbf{p}_{p}^{a}-\mathbf{p}_{l}^{t} \right\|}{c}, \  \tan \left( \theta _{p,l}^{g} \right)  = \frac{y_{l}^{t}-y_{p}^{a}}{ x_{l}^{t}-x_{p}^{a} }.
% 	\end{align}
	where $\theta_{p,l}^{g}$ is the AoA in the global coordinate.
Thus, the Jacobian matrix is expressed as 
$
\mathbf{\Theta }=\frac{\partial \mathbf{\Xi }_l}{\partial \mathbf{p}_{l}^{t}}   =\left[ \begin{array}{cccccc}
	\frac{\partial \tau _{1,l}}{\partial x_{l}^{t}}&		\cdots&		\frac{\partial \tau _{P,l}}{\partial x_{l}^{t}}&		\frac{\partial \psi_{1,l}^{g}}{\partial x_{l}^{t}}&		\cdots&		\frac{\partial \psi_{P,l}^{g}}{\partial x_{l}^{t}}\\
	\frac{\partial \tau _{1,l}}{\partial y_{l}^{t}}&		\cdots&		\frac{\partial \tau _{P,l}}{\partial y_{l}^{t}}&		\frac{\partial \psi_{1,l}^{g}}{\partial y_{l}^{t}}&		\cdots&		\frac{\partial \psi_{P,l}^{g}}{\partial y_{l}^{t}}\\
\end{array} \right],
$
% \begin{equation}
% \mathbf{\Theta }=\frac{\partial \mathbf{\Xi }_l}{\partial \mathbf{p}_{l}^{t}}   =\left[ \begin{array}{cccccc}
% 	\frac{\partial \tau _{1,l}}{\partial x_{l}^{t}}&		\cdots&		\frac{\partial \tau _{P,l}}{\partial x_{l}^{t}}&		\frac{\partial \psi_{1,l}^{g}}{\partial x_{l}^{t}}&		\cdots&		\frac{\partial \psi_{P,l}^{g}}{\partial x_{l}^{t}}\\
% 	\frac{\partial \tau _{1,l}}{\partial y_{l}^{t}}&		\cdots&		\frac{\partial \tau _{P,l}}{\partial y_{l}^{t}}&		\frac{\partial \psi_{1,l}^{g}}{\partial y_{l}^{t}}&		\cdots&		\frac{\partial \psi_{P,l}^{g}}{\partial y_{l}^{t}}\\
% \end{array} \right],
% \end{equation}
where the elements of the Jacobian matrix can be calculated as follows $
\frac{\partial \tau _{p,l}}{\partial x_{l}^{t}}=\frac{2\left( x_{l}^{t}-x_{p}^{a} \right)}{c\sqrt{\left( x_{l}^{t}-x_{p}^{a} \right) ^2+\left( y_{l}^{t}-y_{p}^{a} \right) ^2}},
$
$
\frac{\partial \tau _{p,l}}{\partial y_{l}^{t}}=\frac{2\left( y_{l}^{t}-y_{p}^{a} \right)}{c\sqrt{\left( x_{l}^{t}-x_{p}^{a} \right) ^2+\left( y_{l}^{t}-y_{p}^{a} \right) ^2}},
$
$
\frac{\partial \psi_{p,l}^{g} }{\partial x_{l}^{t}}=\frac{\left( y_{l}^{t}-y_{p}^{a} \right) ^2}{\left( \sqrt{\left( x_{l}^{t}-x_{p}^{a} \right) ^2+\left( y_{l}^{t}-y_{p}^{a} \right) ^2} \right) ^3},
$
$
\frac{\partial \psi_{p,l}^{g} }{\partial y_{l}^{t}}=-\frac{\left( x_{l}^{t}-x_{p}^{a} \right) \left( y_{l}^{t}-y_{p}^{a} \right)}{\left( \sqrt{\left( x_{l}^{t}-x_{p}^{a} \right) ^2+\left( y_{l}^{t}-y_{p}^{a} \right) ^2} \right) ^3}
.$
Finally, the CRLB for estimating the target position $\mathbf{p}_{l}^{t}$ is given by $\mathrm{CRLB}\left( \mathbf{\Xi }_l \right) =\left( \mathbf{\Theta }\mathbf{F}\mathbf{\Theta }^T \right) ^{-1}.$
% \begin{align}
% \mathrm{CRLB}\left( \mathbf{\Xi }_l \right) =\left( \mathbf{\Theta }\mathbf{F}\mathbf{\Theta }^T \right) ^{-1}.
% \end{align}
% \begin{align}
% \frac{\partial \tau _{p,l}}{\partial x_{l}^{t}}=\frac{2\left( x_{l}^{t}-x_{p}^{a} \right)}{c\sqrt{\left( x_{l}^{t}-x_{p}^{a} \right) ^2+\left( y_{l}^{t}-y_{p}^{a} \right) ^2}},
% \\
% \frac{\partial \tau _{p,l}}{\partial y_{l}^{t}}=\frac{2\left( y_{l}^{t}-y_{p}^{a} \right)}{c\sqrt{\left( x_{l}^{t}-x_{p}^{a} \right) ^2+\left( y_{l}^{t}-y_{p}^{a} \right) ^2}},
% \end{align}
% \begin{align}
% \frac{\partial \psi_{p,l}^{g} }{\partial x_{l}^{t}}=\frac{\left( y_{l}^{t}-y_{p}^{a} \right) ^2}{\left( \sqrt{\left( x_{l}^{t}-x_{p}^{a} \right) ^2+\left( y_{l}^{t}-y_{p}^{a} \right) ^2} \right) ^3},
% \\
% \frac{\partial \psi_{p,l}^{g} }{\partial y_{l}^{t}}=-\frac{\left( x_{l}^{t}-x_{p}^{a} \right) \left( y_{l}^{t}-y_{p}^{a} \right)}{\left( \sqrt{\left( x_{l}^{t}-x_{p}^{a} \right) ^2+\left( y_{l}^{t}-y_{p}^{a} \right) ^2} \right) ^3}
% .\end{align}
%\vspace{-2ex}
	\section{Simulation Results}\label{Sec_Simu_Result}
	\begin{figure}[!t]
    \centering
    % --- 第一个子图 (左) ---
    \begin{subfigure}[b]{0.48\linewidth} % 宽度设为 0.48 以留出中间空隙
        \centering
        % 注意：这里的 width=\linewidth 是指相对于子图宽度的 100%
        \includegraphics[width=\linewidth]{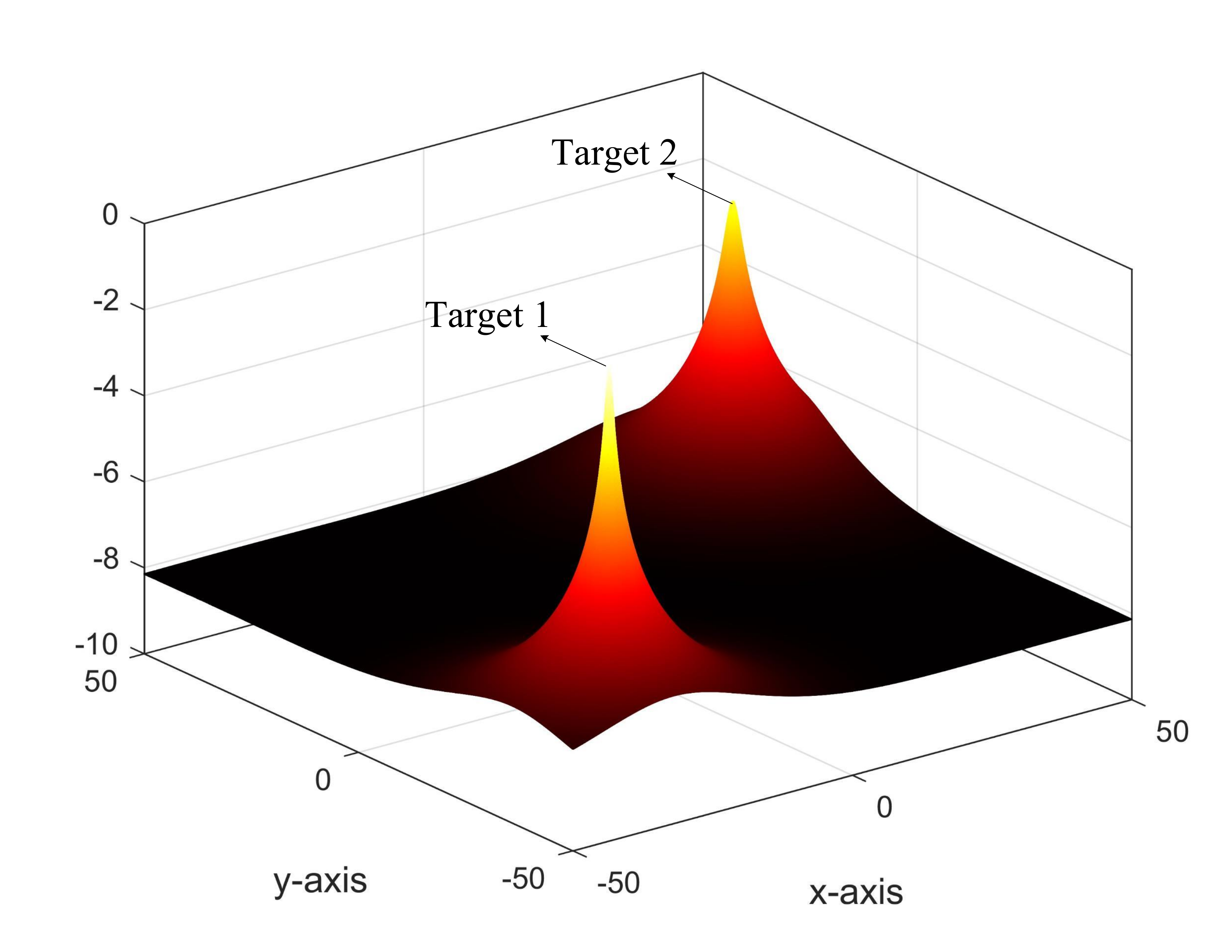}
        \caption{3D plot}
        \label{fig:EVA_s}
    \end{subfigure}
    \hfill % 关键：这个命令把两个图推向两边，形成中间的空隙
    % --- 第二个子图 (右) ---
    \begin{subfigure}[b]{0.48\linewidth}
        \centering
        \includegraphics[width=\linewidth]{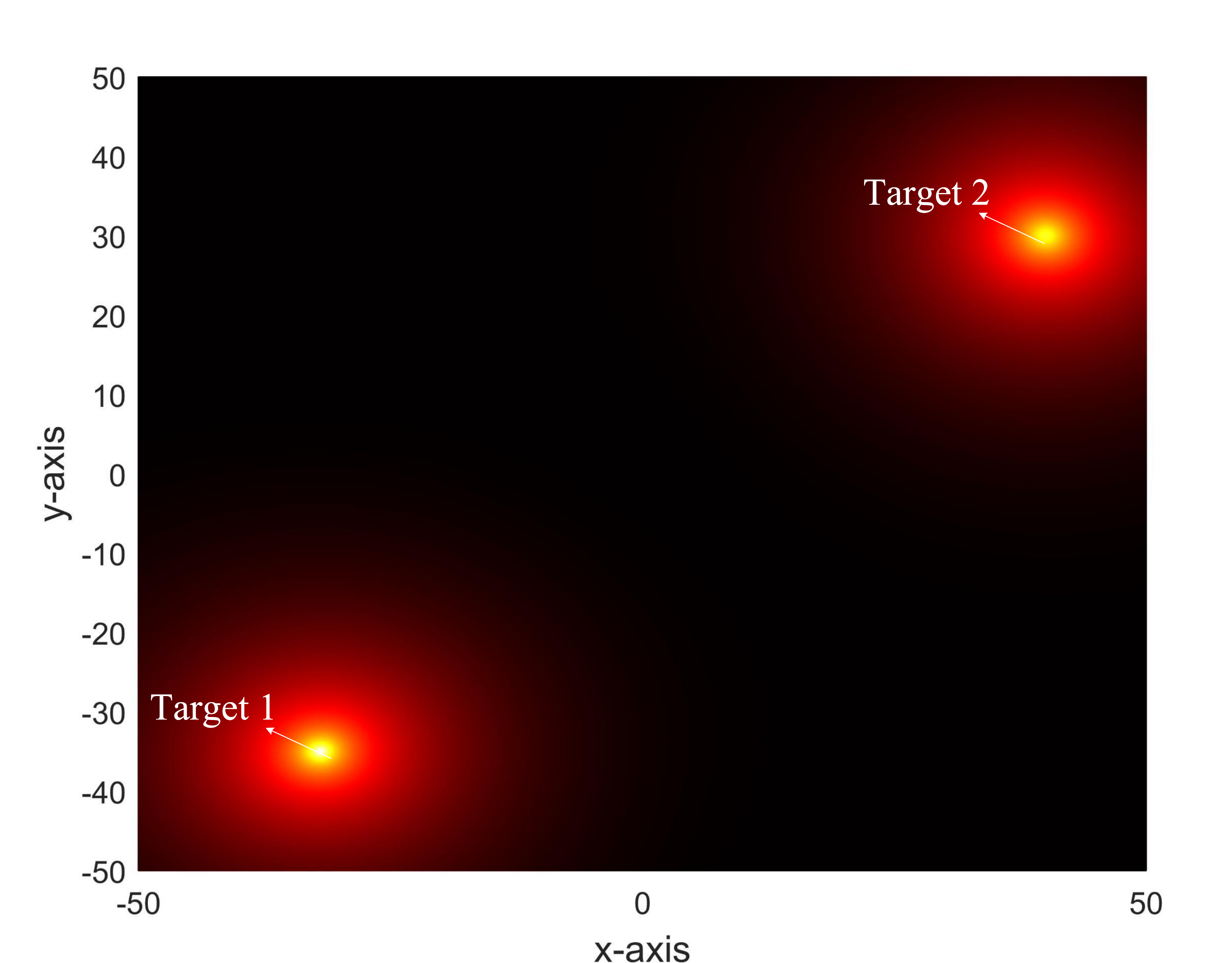}
        \caption{Top view} % 建议修改标题以示区分
        \label{fig:spectrum2.0}
    \end{subfigure}
    % --- 整个大图的标题 ---
    \vspace{-1ex} % 如果需要调整大标题和图的距离
    \caption{Cost function of the proposed algorithm.}
    \label{fig:both_spectra}
\end{figure}
\begin{figure*}[!t]
		\centering
		\vspace{-7ex} % 向上移动
		\includegraphics[width=1 \linewidth]{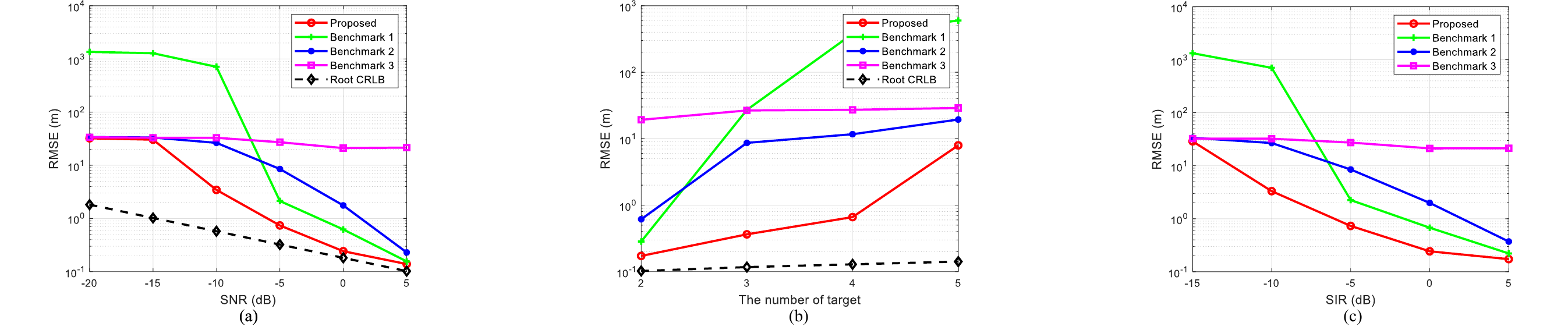}
		\caption{Root CRLB and estimated RMSE versus (a) the SNR, (b) number of targets and \textcolor{blue}{(c) SIR .}}
		\label{fig:simulation}
	\end{figure*}
	Unless noted otherwise, the simulation parameters are set as follows: the system consists of five BSs uniformly distributed along a circle with a 500 m radius. They are equipped with an 8-antenna ULA pointing towards the center of the circle. 
	% The system detects Unmanned Aerial Vehicles (UAVs) as low-altitude targets, assigning each a radar cross-section (RCS) of 0.01 $\text{m}^2$. UAVs are generated uniformly at random within a circular area of radius 400 m and at heights ranging between 50 m and 300 m. They have a maximum flight speed of 60 km/h and a minimum speed of 5 km/h with a safety separation distance of 10 m to avoid collisions. 
	% The radar cross-section (RCS) of each UAV is set to $\sigma _{\mathrm{RCS}} = 0.1$ m\textsuperscript{2}.
	% Sensing signal transmission utilizes a typical sub-6 GHz configuration with a carrier frequency of $f_{\mathrm{c}}=4.9$ GHz and a subcarrier spacing of $\Delta f=30$ kHz. The symbol timing offset and carrier frequency offset are seted as  in the simulations.
	% Unless otherwise stated, each tBS occupies a BWP of 20 MHz (including guard intervals), transmitting sensing signals over 51 resource blocks (RBs), where each RB comprises 12 subcarriers \cite{3GPP13810101}. 
	% Each sensing task employs $N = 7$ consecutive OFDM symbols.
	The system uses a carrier frequency of 4.9 GHz and a subcarrier spacing of $\Delta f=30$ kHz. A sensing task is executed over $N = 7$ OFDM symbols and $8$ resource blocks (RBs). The SNR is set to 5 dB. According to the standard in \cite{3GPP13810101}, each RB contains 12 subcarriers and the total period of OFDM symbol (including the cyclic prefix) is $35.677$ $ \mu s$. The beamforming vectors are randomly selected from the complex unit circle to ensure the generation of an approximately omnidirectional beam \cite{10403776} \cite{10978578}. Two targets with a radar cross-section (RCS) of 0.01 $\text{m}^2$ are located at $[-32, -35]$~m and $[40, 30]$~m, respectively. The path loss is calculated by the following model \cite{ITURRECP525}: 
	$\mathrm{PL}=103.4+20\lg f_{\mathrm{c}}+40\lg d-10\lg \sigma,$
	% \begin{align}
	% \mathrm{PL}=103.4+20\lg f_{\mathrm{c}}+40\lg d-10\lg \sigma,
	% \end{align}
	where $d$ is the distance in km from the target to the AP and $f_c$ is the carrier frequency in MHz and $\sigma$ is the RCS in m\textsuperscript{2}. The simulation results are averaged over 1,000 independent Monte Carlo trials. The sensing performance is evaluated using the root mean square error (RMSE) metric, i.e.,
	$
	\text{RMSE}(\mathbf{p}^t) = \sqrt{\frac{1}{L} \sum_{l=1}^{L} \left\| \hat{\mathbf{p}}_l^t - \mathbf{p}_l^t \right\|^2}.
$
% 	\begin{equation}
% 	\text{RMSE}(\mathbf{p}^t) = \sqrt{\frac{1}{L} \sum_{l=1}^{L} \left\| \hat{\mathbf{p}}_l^t - \mathbf{p}_l^t \right\|^2}.
% \end{equation}
To evaluate the proposed cooperative sensing algorithm, we further compare it with three baselines, defined as follows:
	\begin{itemize}
		\item \textbf{Benchmark 1}: As proposed in \cite{JunTang}, APs first utilize the spatial smoothing-based tensor decomposition to estimate the parameters of the targets. Then, data association and fusion sensing are achieved by solving a minimum spanning tree and a nonlinear optimization problem.
		\item \textbf{Benchmark 2}: Inspired by \cite{zhenkunzhang}, APs utilize only the time delay information for sensing instead of constructing EVA, while its fusion framework follows the subspace-based approach.
		% \item \textbf{Banchmark 3}: APs utilize only the angle information for sensing instead of constructing EVA, while its fusion framework remains based on the subspace algorithm.
		\item \textbf{Benchmark 3}: As proposed in \cite{liu2025bayesianprobabilityfusionmultiap}, APs implement a Bayesian probabilistic fusion algorithm by constructing a constrained maximum a posteriori (MAP) problem.
	\end{itemize}
	%  of the most accurate estimations, discarding the worst-performing 5%.
	
	% \begin{align}
	% 	&\mathrm{PL}_{\mathrm{Ts}}=103.4+20\lg f_{\mathrm{c}}+20\lg d^{\mathrm{t}}+20\lg d^{\mathrm{r}}-10\lg \sigma _{\mathrm{RCS}},
	% 	\\
	% 	&\mathrm{PL}_{\mathrm{LoS}}=32.4+20\lg f_{\mathrm{c}}+20\lg d^{\mathrm{LoS}},
	% \end{align}
	% where $d^{\mathrm{t}}$ and $d^{\mathrm{r}}$ are the distances in km from the target to the tBS and rBS, respectively, and $d^{\mathrm{LoS}}$ denotes the distance in km between the tBS and rBS.
	% The root mean square error (RMSE) is employed to quantify the sensing accuracy in the simulations. 
	% For any sensing quantity $\mathbf{x}$, which may represent bistatic ranges, Doppler velocities, AoA estimates, or 3D position/velocity estimates, the RMSE is defined as:
	% \begin{equation}
	% 	\mathrm{RMSE}\left( \mathbf{x} \right) = \sqrt{\frac{1}{K}\sum_{k=1}^K{\left\| \hat{\mathbf{x}}_k - \mathbf{x}_k \right\| _{2}^{2}}},
	% \end{equation}
	% where $\hat{\mathbf{x}}_k$ and $\mathbf{x}_k$ denote the estimated and true value for target $k$, respectively.
	% All simulation results are derived from 500 independent Monte Carlo trials.
	% To mitigate the influence of outliers, the reported results represent the average over the top 95\% of samples with the highest estimation accuracy, excluding the bottom 5\%.
	% Note that all simulation results are presented on a logarithmic scale for the y-axis.
	
	%\subsection{Cooperative Localization Results}

Fig. \ref{fig:EVA_s} and Fig. \ref{fig:spectrum2.0} intuitively demonstrate that the cost function yields two distinct peaks precisely at the true target locations. The step-by-step procedure for searching these spectral peaks is outlined in Algorithm~\ref{algorithm:subspace}. 

As shown in Fig. \ref{fig:simulation}a, the proposed algorithm achieves the best performance. The localization error of the proposed algorithm is significantly lower than that of the existing benchmarks and closely approaches the square root of the CRLB (Root CRLB). 
% \textcolor{blue}{This superiority stems from the EVA-enabled association-free subspace fusion, which avoids data association and directly exploits cooperative gain across APs.} 
\textcolor{blue}{This superiority stems from the EVA-enabled association-free subspace fusion, which preserves richer spatial-frequency observation information and exploits the cooperative gain from a subspace perspective, thereby avoiding explicit data association among local estimates.} 
In contrast, the data association in Benchmark 1 relies on the parameter extraction results from tensor decomposition of the single AP, which leads to severe performance degradation when parameter extraction fails. The performance of Benchmark 2 not only validates the feasibility of the subspace fusion method but also confirms that constructing the EVA array effectively improves sensing accuracy. Benchmark~3 is restricted by its inherent error floor because the Bayesian probability-based fusion method exhibits limited performance in multi-target scenarios.

Fig. \ref{fig:simulation}b shows the RMSE versus the number of targets. It is evident that the proposed algorithm exhibits superior scalability, maintaining a much lower RMSE compared to other benchmarks. Specifically, the scalability of Benchmark 1 is limited by the fragility of its single-AP processing, which results in frequent association mismatches as the number of targets increases.
 Alternatively, Benchmark 3 suffers from severe resolution loss. The proposed algorithm mitigates this drawback via subspace fusion. The performance of Benchmark 2 is limited by the number of subcarriers. The proposed algorithm mitigates this drawback via constructing the EVA array to expand the sensing aperture.

\textcolor{blue}{Fig. \ref{fig:simulation}c evaluates the RMSE performance versus the signal-to-interference ratio (SIR), defined as  $(\|\mathcalbf{Y}_p-\mathcalbf{N}_p\|_F^2)/
 \|\mathcalbf{I}_p\|_F^2$. The element of interference tensor $\mathcalbf{I}_p$ is modeled as an independent complex Gaussian variable, which may originate from imperfect signal separation even under frequency-division schemes, in-band transmissions from other devices, or AP residual self-interference \cite{wang2015outage}. The proposed algorithm consistently outperforms the benchmarks, indicating its robustness against interference. This advantage is attributed to the EVA construction and association-free multi-AP subspace fusion, which jointly exploit spatial, frequency and multi-AP domains to alleviate interference-induced subspace perturbations without suffering from association errors.}

	\vspace{-1.2ex}
	\section{Conclusion}\label{Sec_Conclusion}
	In this paper, we proposed a novel data association-free subspace fusion algorithm for cooperative ISAC systems. We constructed an EVA array to expand the sensing aperture via tensor unfolding, which effectively bypasses the complex data association procedures. Furthermore, a derivation of the CRLB was presented. Simulation results demonstrated that the proposed algorithm achieved the best sensing performance compared with existing techniques and closely approached the Root CRLB.
	\bibliographystyle{IEEEtran}
	\bibliography{IEEEabrv,ref}

@STRING{IEEE_J_JSAC       = "{IEEE} J. Sel. Areas Commun."}

@STRING{IEEE_J_WCOM       = "{IEEE} Trans. Wireless Commun."}

@article{gill1972quasi,
  title={Quasi-Newton methods for unconstrained optimization},
  author={Gill, Philip E and Murray, Walter},
  journal={IMA J. Appl. Math.},
  volume={9},
  number={1},
  month = apr,
  pages={91--108},
  year={1972},
  publisher={Oxford University Press}
}

@ARTICLE{9737357,
  author={Liu, Fan and others},
  journal=IEEE_J_JSAC, 
  title={Integrated Sensing and Communications: Toward Dual-Functional Wireless Networks for 6{G} and Beyond}, 
  year={2022},
  month = oct,
  volume={40},
  number={6},
  pages={1728-1767},
  doi={10.1109/JSAC.2022.3156632}}

@ARTICLE{JunTang,
 	author={Tang, Jun and others},
 	journal={IEEE Trans. Wireless Commun.}, 
 	title={Cooperative {ISAC}-Empowered Low-Altitude Economy}, 
 	year={2025},
  month = may,
 	volume={24},
 	number={5},
 	pages={3837-3853},
 }

@ARTICLE{zhenkunzhang,
  author={Zhang, Zhenkun and others},
  journal={IEEE Trans. Commun.}, 
  title={Target Localization in Cooperative {ISAC} Systems: A Scheme Based on {5G} {NR} {OFDM} Signals}, 
  year={2025},
  volume={73},
  number={5},
  pages={3562 -- 3578},
  month = may,
}

@ARTICLE{liu2025bayesianprobabilityfusionmultiap,
      title={Bayesian Probability Fusion for Multi-AP Collaborative Sensing in Mobile Networks}, 
      author={Shengheng Liu and others},
      year={2025},
      journal={arXiv:2512.02462},
}

@article{zhang2025modular,
  title={Modular {XL}-Array-Enabled {3-D} Localization based on Hybrid Spherical-Planar Wave Model in Terahertz Systems},
  author={Zhang, Yang and others},
  journal={arXiv preprint arXiv:2504.13455},
  year={2025}
}

@ARTICLE{10403776,
  author={Zhang, Ruoyu and others},
  journal=IEEE_J_WCOM,
  title={Integrated Sensing and Communication With Massive {MIMO}: A Unified Tensor Approach for Channel and Target Parameter Estimation}, 
  year={2024},
  month = aug,
  volume={23},
  number={8},
  pages={8571-8587},
  doi={10.1109/TWC.2024.3351856}}

@INPROCEEDINGS{10978578,
  author={Liu, Haotian and others},
  booktitle={Proc. IEEE Wireless Commun. Netw. Conf.
(WCNC)}, 
  title={Multipath Component-Aided Signal Processing for Integrated Sensing and Communication Systems}, 
  year={2025},
  volume={},
  number={},
  pages={1-7},
  doi={10.1109/WCNC61545.2025.10978578}}

@techreport{3GPP13810101, 
	author = {3GPP}, 
	institution = {{3rd Generation Partnership Project (3GPP)}}, 
	month = jul, 
	day={25},
	note = {Version 17.10.0}, 
	number = {38.101-1}, 
	title = {{5G}; {NR}; User Equipment ({UE}) radio transmission and reception; Part 1: Range 1 Stand alone}, 
	type = {Technical Specification (TS)}, 
	url = {	https://www.etsi.org/deliver/etsi_ts/138100_138199/13810101/.}, 
	year = {2023}}

@techreport{ITURRECP525, 
	author = {ITU}, 
	institution = {{International Telecommunication Union (ITU)}}, 
	month = aug, 
	day={11},
	note = {}, 
	number = {{P.525-4}}, 
	title = {Calculation of free-space attenuation }, 
	type = {Recommendation (R)}, 
	url = {https://www.itu.int/rec/R-REC-P.525.}, 
	year = {2019}}

@ARTICLE{10870338,
  author={Xiu, Yue and others},
  journal={IEEE Wireless Commun. Lett.}, 
  title={Movable Antenna Enabled {ISAC} Beamforming Design for Low-Altitude Airborne Vehicles}, 
  	month = feb, 
  year={2025},
  volume={14},
  number={5},
  pages={1311-1315},
  doi={10.1109/LWC.2025.3538587}}

@article{shi2022device,
  author  = {Shi, Qin and Liu, Liang and Zhang, Shuowen and Cui, Shuguang},
  title   = {{Device-Free Sensing in OFDM Cellular Network}},
  journal = {IEEE J. Sel. Areas Commun.},
  volume  = {40},
  number  = {6},
  pages   = {1838--1853},
  year    = {2022},
  month   = jun,
  doi     = {10.1109/JSAC.2022.3155543}
}

@ARTICLE{wang2015outage,
  author  = {Q. Wang and Y. Dong and X. Xu and X. Tao},
  journal = {IEEE Commun. Lett.},
  title   = {Outage probability of full-duplex {AF} relaying with processing delay and residual self-interference},
  year    = {2015},
  volume  = {19},
  number  = {5},
  month = mar,
  pages   = {783--786},
  doi     = {10.1109/LCOMM.2015.2411596}
}

\end{document}